\documentclass[twocolumn,abs,prb,showpacs,floatfix]{revtex4-1}

\usepackage{graphicx}
\usepackage{dcolumn}
\usepackage{float}
\usepackage{amsmath}
\usepackage{bm}
\usepackage{natbib}
\usepackage[mathscr]{euscript}

\begin{document}


\title[]{Magnetization of the Metallic Surface States in Topological Insulators}

\author{C. J. Tabert$^{1,2}$\footnote{Corresponding author: ctabert@uoguelph.ca} and J. P. Carbotte$^{3,4}$}
\address{$^1$Department of Physics, University of Guelph,
Guelph, Ontario N1G 2W1 Canada} 
\address{$^2$Guelph-Waterloo Physics Institute, University of Guelph, Guelph, Ontario N1G 2W1 Canada}
\address{$^3$Department of Physics, McMaster University,
Hamilton, Ontario L8S 4M1 Canada} 
\address{$^4$Canadian Institute for Advanced Research, Toronto, Ontario M5G 1Z8 Canada}
\date{\today}

\begin{abstract}
{We calculate the magnetization of the helical metallic surface states of a topological insulator.  We account for the presence of a small sub-dominant Schr{\"o}dinger piece in the Hamiltonian in addition to the dominant Dirac contribution.  This breaks particle-hole symmetry.  The cross-section of the upper Dirac cone narrows while that of the lower cone broadens.  The sawtooth pattern seen in the magnetization of the pure Dirac limit as a function of chemical potential ($\mu$) is shifted; but, the quantization of the Hall plateaus remains half integral.  This is verified by taking the derivative of the magnetization with respect to $\mu$.  We compare our results with those when the non-relativistic piece dominates over the relativistic contribution and the quantization is integral.  Analytic results for the magnetic oscillations are obtained where we include a first order correction in the ratio of non-relativistic to relativistic magnetic energy scales.  Our fully quantum mechanical derivations confirm the expectation of semiclassical theory except for a small correction to the expected phase.  There is a change in the overall amplitude of the magnetic oscillations.  The Dingle and temperature factors are modified.
}
\end{abstract}


\maketitle

\section{Introduction}
A topological insulator is a bulk insulator with a metallic spectrum of topologically protected helical surface states\cite{Kane:2005, Hasan:2010, Qi:2011}.  The helical fermions which exist at the surface\cite{Chen:2009, Hsieh:2009, Chen:2010, Xu:2011} display an odd number of Dirac points.  As examples, Bi$_2$Se$_3$\cite{Hsieh:2009} has a single point while samarium hexaboride has three\cite{Roy:2014}.  In such systems, the in-plane spin of the electron is perpendicular to its momentum.  Recently, angle-resolved photoemission spectroscopy (ARPES) experiments have mapped out the surface bands\cite{Chen:2009, Hsieh:2009} and confirmed the predicted Dirac-like spectrum and spin arrangement.  In contrast to graphene, whose low-energy Dirac cones are particle-hole symmetric, the surface states of a topological insulator exhibit band bending and take an hourglass shape with the valence band below the Dirac point displaying significantly more outward bending than the corresponding inward-bending of the conduction band\cite{Chen:2009, Hsieh:2009, Xu:2011, Hancock:2011, Wright:2013, ZLi:2013a, ZLi:2013}.  This behaviour can be captured by adding a Schr{\"o}dinger mass term to the ideal linear Dirac Hamiltonian which remains dominant in topological insulators.  The resulting particle-hole asymmetry has important ramifications on the physics of such systems.  As examples, Wright and Mackenzie\cite{Wright:2013} have discussed its effect on the Berry phase while Wright\cite{Wright:2013a} describes its  role on measurements of the Chern number and the phase transition from the spin Hall to quantum anomalous Hall phase.  Shubnikov-de-Haas (SdH) oscillations\cite{Fuchs:2010} are also expected to be altered\cite{Taskin:2011a, Mikitik:2012, Ando:2013, Raoux:2014, Kishigi:2014}. Li \emph{et al}\cite{ZLi:2014} have considered the Hall conductivity and optical absorption\cite{ZLi:2013}.  Particle-hole asymmetry splits the interband magneto-optical absorption lines of the pure Dirac case\cite{Gusynin:2007, Pound:2012} into two.  This is due to the broken degeneracy of the energy difference between the valence band Landau levels $-m$ and $-(m+1)$ to conduction levels $m+1$ and $m$, respectively.  Schafgans \emph{et al}\cite{Schafgans:2012} have given results of magneto-optical measurements in the topological insulator Bi$_{0.91}$Sb$_{0.09}$.

Here we consider the magnetization ($M$) and, in particular, describe how the sub-dominant Schr{\"o}dinger term changes its dependence on chemical potential ($\mu$).  The derivative of $M$ with respect to $\mu$ is of particular interest as it is related to the underlying quantization of the Hall plateaus.  We compare our results with those in the opposite limit when the Schr{\"o}dinger term dominates.  This yields a nearly parabolic electronic dispersion which is slightly modified by a small spin-orbit interaction\cite{Bychkov:1984a, Bychkov:1984}.  This limit is relevant to the field of spintronic semiconductors which has been extensively studied in the past\cite{Bychkov:1984a, Bychkov:1984,Luo:1990, deAndrada:1994, Nitta:1997, Grundler:2000, Wang:2005, Zarea:2005,Zutic:2004, Fabian:2007}.   In addition to the Hall plateaus, the oscillations in the magnetization are also affected by the Schr{\"o}dinger term.

Our manuscript is organized as follows: In Sec.~II, we introduce the appropriate low-energy Hamiltonian and Landau level spectrum resulting from a finite magnetic field.  Section~III contains a discussion of the grand thermodynamic potential on which our calculations are based.  Both relativistic and non-relativistic limits are considered.  Numerical results are presented for the evolution of the magnetization as a function of chemical potential for several ratios of the relevant energy scales.   The results are compared to the pure Dirac limit and the differences are emphasized.  We calculate the derivative $[\partial M(\mu)/\partial \mu]\vline_B$, which is related to the quantization of the Hall conductivity through the Streda formula\cite{Wang:2010}.  For comparison, similar results are obtained in the spintronic regime where the Schr{\"o}dinger scale dominates and the Dirac term is a small perturbation.  In Sec.~IV, we give a fully quantum mechanical derivation of the effect a sub-dominant Schr{\"o}dinger term has on the quantum oscillations.  Finite temperature effects are discussed as is the effect of impurity scattering in the constant scattering rate approximation.  Our conclusions follow in Sec.~V.

\section{Low-Energy Hamiltonian}

\begin{figure}[h!]
\begin{center}
\includegraphics[width=1\linewidth]{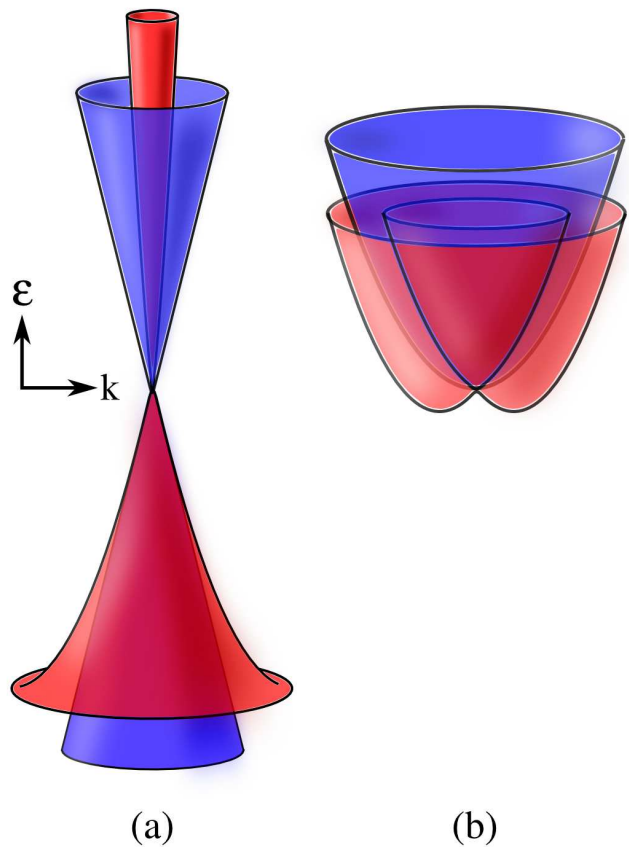}
\end{center}
\caption{\label{fig:Energy}(Color online) (a) Pure-Dirac cones (blue) overlaid with those of a topological insulator (red) where a small Schr{\"o}dinger mass term has been included.  With the mass term, the conduction band narrows while the valence band fans out.  (b) Pure-Schr{\"o}dinger dispersion (blue) overlaid with the band structure of a spintronic semiconductor (red) in which a small $v_F$ term has been included.  The resulting band structure resembles two offset parabolic bands.
}
\end{figure}

In the absence of a magnetic field, the low-energy helical surface fermions of a topological insulator are well described by the Bychkov-Rashba Hamiltonian\cite{Bychkov:1984a, Bychkov:1984}
\begin{align}\label{HAM}
H=\frac{\hbar^2 k^2}{2m}+\hbar v_F (k_x\sigma_y-k_y\sigma_x),
\end{align}
where $\sigma_x$ and $\sigma_y$ are the usual Pauli matrices associated with spin and $\bm{k}$ is the momentum measured relative to the $\Gamma$ point of the surface Brillouin zone.  The first term is the familiar parabolic Schr{\"o}dinger piece for describing an electron with effective mass $m$.  The second term describes massless Dirac fermions which move with a Fermi velocity $v_F$.  Equation~\eqref{HAM} can be solved to give the energy dispersion
\begin{align}
\varepsilon_\pm(\bm{k})=\frac{\hbar^2k^2}{2m}\pm\hbar v_F k.
\end{align}
A schematic of the energy dispersion is shown in Fig.~\ref{fig:Energy}.
Figure~\ref{fig:Energy}(a) shows the surface-state band structure in the topological insulator regime (i.e. the Dirac term dominates).  The blue cones correspond to the pure Dirac limit ($m\rightarrow\infty$) while the red dispersion results from a non-infinite mass.  In Fig.~\ref{fig:Energy}(b), the spintronic limit is shown (i.e. the Schr{\"o}dinger term dominates).  The blue parabola corresponds to the pure Schr{\"o}dinger case ($v_F\rightarrow 0$) while the two offset red parabolas result from a finite $v_F$.  It should be clear from this figure that, while the same Hamiltonian describes both sets of dispersion curves, in the confines of a Brillouin zone, these are very different and, as will be seen, lead to different physics.  In fact, the low-energy Hamiltonian does not allow these two regimes to be connected continuously; but, it does describe each separately given appropriate values of the two characterizing parameters $v_F$ and $m$.  For the specific case of the topological insulator Bi$_2$Te$_3$, detailed band structure calculations\cite{Zhang:2009,Liu:2010} give: $v_F=4.3\times 10^5$m/s, and $m=0.09m_e$ where $m_e$ is the bare mass of an electron\cite{ZLi:2014}.  For the typical spintronic materials, $v_F$ is reduced to $\mathcal{O}(10^3)$m/s at most\cite{Fabian:2007, Zutic:2004} and $m$ remains at approximately 0.09$m_e$.  The continuum Hamiltonian of Eqn.~\eqref{HAM} allows the valence band to bend back over and eventually cross the zero energy axis.  This is not physical in the context of topological insulators.  Therefore, when using this model, it is important to apply an appropriate momentum cutoff to prevent this spurious behaviour.

To discuss the theory of magnetic oscillations, we need to consider the effect of an external magnetic field $B$ which we orient perpendicular to the surface of the insulator ($\hat{z}$).  To do this, we work in the Landau gauge so that the magnetic vector potential, given by $\bm{B}=\nabla\times\bm{A}$, is written as $\bm{A}=(0,Bx,0)$.  We then make the Peierls substitution on the momentum $\hat{p}_i\rightarrow\hat{p}_i-q\hat{A}_i$, where $q=-e$ is the charge of the electron.  Thus, Eqn.~\eqref{HAM} becomes
\begin{align}\label{HAM-B}
 H=&\frac{\hbar^2}{2m}\left[(-i\partial_x)^2+(-i\partial_y+eBx/\hbar)^2\right]\notag\\
 &+\hbar v_F[(-i\partial_x)\sigma_y-(-i\partial_y+eBx/\hbar)\sigma_x].
\end{align}
Here, we are not including a Zeeman splitting term.  Its possible effects on the magnetization have already been considered in the work of Wang \emph{et al}\cite{Wang:2010} who found it to have a negligible effect on all the Landau level energies except for that at $N=0$.  Therefore, it will not appreciably affect the phenomena of interest in this paper. Eqn.~\eqref{HAM-B} can be solved to give the Landau level dispersion
\begin{align}
E_{N,s}=\frac{\hbar^2N}{ml_B^2}+s\sqrt{\left(\frac{\hbar^2}{2ml_B^2}\right)^2+\frac{2N\hbar^2v_F^2}{l_B^2}},
\end{align}
where $l_B=\sqrt{\hbar/(e|B|)}$ is the magnetic coherence length, $s=\pm$ for the conduction and valence bands, respectively, and $N> 0$ is an integer and gives the Landau level index.  The $N=0$ Landau level must be treated carefully and is given by
\begin{align}
E_{N=0}=\frac{\hbar^2}{2ml_B^2}.
\end{align}
For convenience, we will define both Schr{\"o}dinger and Dirac energy scales which are given by $E_0=\hbar e|B|/m$ and $E_1=\hbar v_F\sqrt{e|B|/\hbar}$, respectively.  It is also useful to introduce a dimensionless parameter $P=E_1^2/E_0^2$.  The limit $P\rightarrow\infty$ corresponds to the pure Dirac system while $P\rightarrow 0$ in the pure Schr{\"o}dinger case.

Using these definitions, the Landau level spectrum for $N\neq 0$ can be expressed as
\begin{align}
E_{N,s}=E_0N+s\sqrt{\left(\frac{E_0}{2}\right)^2+2NE_1^2}
\end{align}
or, equivalently,
\begin{align}
E_{N,s}=E_0N+sE_0\sqrt{\frac{1}{4}+2NP}.
\end{align}
For $N=0$,
\begin{align}
E_{N=0}=\frac{E_0}{2}.
\end{align}

One must be careful when dealing with the $s=-$ levels for $E_1>E_0$. As previously mentioned, we are working with a continuum model in which the valence band can bend back toward the zero energy axis.  As a result, that unphysical portion of the band structure can also condense into Landau levels.  Indeed, one finds that for large $N$, the $E_{N-}$ levels begin to increase in energy.  While some bending may be characteristic of a topological insulator, one must not allow the valence band to cross the energy axis.  This is done by applying a momentum cutoff when $B=0$.  For a finite $B$, we must apply an appropriate cutoff on $N$ to ensure none of the $s=-$ levels become positive.

\section{Grand Thermodynamic Potential}

\subsection{Dirac limit}

\begin{figure}[h!]
\begin{center}
\includegraphics[width=0.9\linewidth]{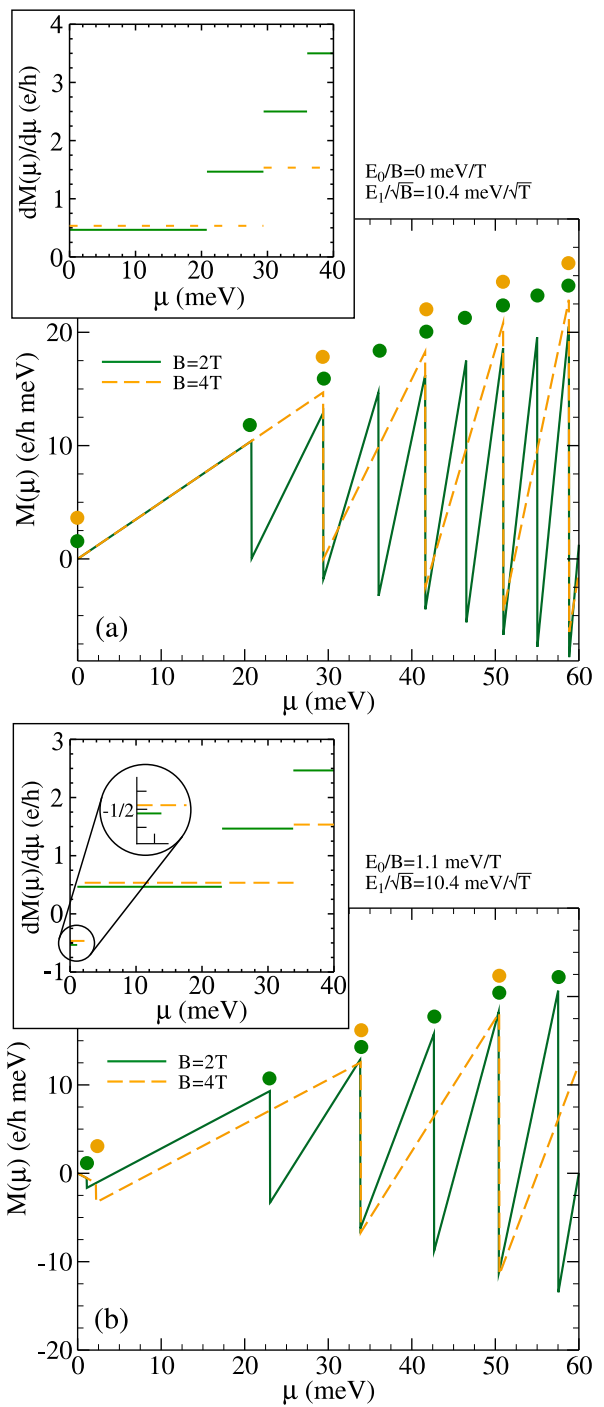}
\end{center}
\caption{\label{fig:Mag-Dirac-m}(Color online) Magnetization as a function of $\mu$ for the (a) pure-Dirac limit and (b) topological insulator regime ($E_0\ll E_1$).  The results are shown for two fixed values of magnetic field.  The circles above the peaks illustrate the intersection of the constant $B$ lines with the Landau levels (see Fig.~\ref{fig:LL-Dirac}).  The insets show $(h/e)\partial M(\mu)/\partial\mu=(h/e^2)\sigma_H$.  In both cases, $\sigma_H$ is quantized in half-integer values of $e^2/h$.  Note: the Hall plateaus have been slightly offset from the half-integer values for clarity.
}
\end{figure}

\begin{figure}[h!]
\begin{center}
\includegraphics[width=0.9\linewidth]{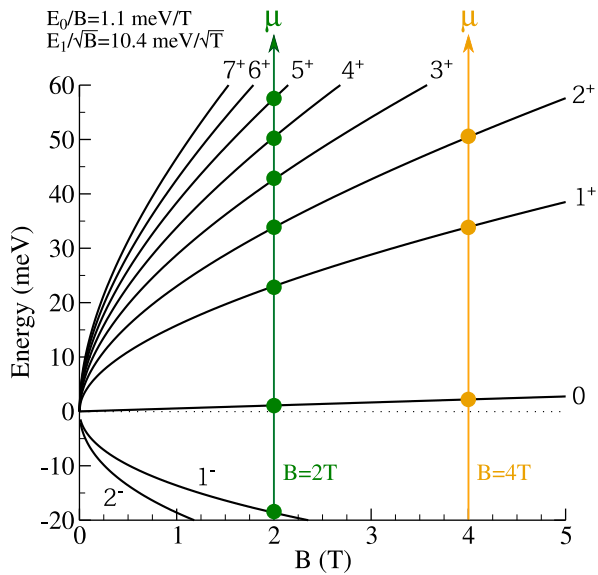}
\end{center}
\caption{\label{fig:LL-Dirac}(Color online) Landau level dispersion as a function of $B$ for the topological insulator regime ($E_0\ll E_1$).  The Landau level index is given by the numbers near the curves; the superscripts $\pm$ give the band index.  Lines are drawn to correspond to the constant $B$ values used in Fig.~\ref{fig:Mag-Dirac-m}.  The intersections of these lines with the Landau levels are illustrated by the circles.
}
\end{figure}

Our discussion of the magnetic response of the surface charge carriers begins with the grand thermodynamic potential $\Omega(T,\mu)$.  For the relativistic Dirac case with particle-hole symmetry, Sharapov \emph{et al}\cite{Sharapov:2004} start from
\begin{align}\label{Omega}
\Omega(T,\mu)=-T\int_{-\infty}^{\infty}N(\omega) \rm{ln}\left(2\rm{cosh}\frac{\omega-\mu}{2T}\right)d\omega ,
\end{align}
where $T$ is the temperature, $\mu$ is the chemical potential, and $N(\omega)$ is the density of states.  In the absence of impurities, $N(\omega)$ is a series of Dirac delta functions located at the various Landau level energies.  Equation~\eqref{Omega} can be rewritten as
\begin{align}\label{Omega2}
\Omega(T,\mu)&=-T\int_{-\infty}^{\infty}N(\omega)\rm{ln}\left(1+e^{(\mu-\omega)/T}\right)d\omega\notag\\
&+\frac{1}{2}\int_{-\infty}^{\infty}\mu N(\omega)d\omega -\frac{1}{2}\int_{-\infty}^{\infty}\omega N(\omega)d\omega .
\end{align}
The first term on the right hand side has the form of the usual non-relativistic grand potential [$\Omega_{\rm NR}(T,\mu)$]; except, now there are negative energy states.  The second term is half the chemical potential times the total number of states in our bands.  Therefore, it does not contribute to the magnetization [$M(T,\mu)$] which is given by the first derivative of $\Omega(T,\mu)$ with respect to $B$ at fixed chemical potential, i.e., $M(T,\mu)=-[\partial\Omega(T,\mu)/\partial B]\vline_\mu$.   The final term is zero in graphene because of particle-hole symmetry.  Thus, the magnetization calculated from the relativistic grand potential reduces correctly to that of $\Omega_{\rm NR}$.  While we could proceed directly from $\Omega_{\rm NR}$, it is convenient to keep the second term of Eqn.~\eqref{Omega2}.  At zero temperature, the first two terms of Eqn.~\eqref{Omega2} reduce to
\begin{align}\label{Omega-T0}
\Omega(T=0,\mu)&=\frac{1}{2}\int_{-\infty}^{0^-}(2\omega-\mu)N(\omega)d\omega\notag\\
&+\int_{0^+}^{\mu}(\omega-\mu)N(\omega)d\omega+\frac{1}{2}\int_{0^+}^{\infty}\mu N(\omega)d\omega ,
\end{align}
which can be used to derive the results for graphene as well as a topological insulator.  The density of states for our topological insulator has the form
\begin{align}\label{DOS}
N(\omega)=\frac{eB}{h}\left[\delta\left(\omega-E_0/2\right)+\sum_{N=1, s=\pm}^\infty\delta(\omega-E_{N,s})\right],
\end{align}
where $s=\pm$ gives the conduction and valence band, respectively.  For graphene, $E_0=0$ and hence the first term in Eqn.~\eqref{DOS} is a Dirac delta function at $\omega=0$ which must be duly noted.  For graphene, $\int_{0^+}^\infty N(\omega)d\omega=\int_{-\infty}^{0^-} N(\omega)d\omega$ while for a topological insulator with all $E_{N,-}<0$, $\int_{0^+}^\infty N(\omega)d\omega=\int_{-\infty}^{0^-} N(\omega)d\omega+eB/h$ and thus 
\begin{align}\label{Omega-int}
\Omega(0,\mu)=\int_{0^+}^\mu (\omega-\mu) N(\omega)d\omega +\int_{-\infty}^{0^-}\omega N(\omega)d\omega+\frac{eB\mu}{2h},
\end{align}
for a topological insulator, and
\begin{align}
\Omega(0,\mu)=\int_{0^+}^\mu (\omega-\mu)N(\omega)d\omega +\int_{-\infty}^{0^-}\omega N(\omega)d\omega,
\end{align}
for graphene.  In both cases, the vacuum contribution $\Omega_0(0)$ is $\int_{-\infty}^{0^-}\omega N(\omega)d\omega$ which can depend on $B$ but not on $\mu$.  For graphene, this has been worked out in detail by Sharapov \emph{et al}\cite{Sharapov:2004} and found to go like $B^{3/2}$ [see their Eqn.~(A5)].  When interested in the changes in magnetization for fixed $B$ as a function of $\mu$, this term can be dropped as it simply adds a constant background.  We also note that
\begin{align}
\left(\frac{\partial\Omega(0,\mu)}{\partial \mu}\right)_B=-\int_{0^+}^\mu N(\omega)d\omega+\frac{eB}{2h},
\end{align}
for a topological insulator,
\begin{align}
\left(\frac{\partial\Omega(0,\mu)}{\partial \mu}\right)_B=-\int_{0^+}^\mu N(\omega)d\omega,
\end{align}
for graphene, and that the vacuum does not appear in this quantity.  For a topological insulator excluding the vacuum contribution, we obtain,
\begin{align}\label{Omega-TI}
\tilde{\Omega}(0,\mu)&=\frac{eB}{h}\left[\frac{\mu}{2}+\left(E_0/2-\mu\right)\Theta\left(\mu-E_0/2\right)\right.\notag\\
&\left.+\sum_{N=1}^\infty\left(E_{N,+}-\mu\right)\Theta\left(\mu-E_{N,+}\right)\right],
\end{align}
where we have used Eqn.~\eqref{DOS} for the density of states, and have assumed that all $E_{N,-}$ energies remain negative.  The magnetization $M(\mu)$ as a function of $\mu$ derived from Eqn.~\eqref{Omega-TI} is, by arrangement, zero for zero chemical potential.  In Fig.~\ref{fig:Mag-Dirac-m}, we display results for $M(\mu)$ as a function of $\mu$ for two fixed values of magnetic field ($B=2$ and 4 Tesla). 
Including a vacuum contribution would simply shift the curves up by a constant.  In all cases, the Dirac energy $E_1/\sqrt{B}$ is set at 10.4 meV/$\sqrt{\rm T}$ which is a characteristic value of topological insulators.  Two values of the Schr{\"o}dinger scale are considered.  $E_0/B=0$ [Fig.~\ref{fig:Mag-Dirac-m}(a)] corresponds to the pure Dirac case (ex. graphene) and is included only for comparison.  Figure~\ref{fig:Mag-Dirac-m}(b) shows $E_0/B=1.1$meV/T.  In all cases, the magnetization displays a saw-tooth oscillation pattern where the location of the vertical jumps is best understood by examining a plot of the Landau level dispersion shown in Fig.~\ref{fig:LL-Dirac} for the specific case of $E_0/B=1.1$meV/T.
The black lines correspond to the Landau level energies as a function of $B$.  The vertical green and orange lines correspond to the two values of $B$ which yield the curves in Fig.~\ref{fig:Mag-Dirac-m}(b).  The colored circles mark the intersection of the Landau levels with the lines of constant $B$.  It is clear that the intersection of the Landau levels with these constant magnetic field lines, correspond to the vertical jumps in $M(\mu)$ seen in Fig.~\ref{fig:Mag-Dirac-m}(b).

The slope of $M(\mu)$ is of particular importance as it is related to the quantized Hall conductivity ($\sigma_H$) through the Streda formula\cite{Wang:2010} $\partial M(\mu)/\partial\mu=(1/e)\sigma_H$.  Indeed, by examining the inset of Fig.~\ref{fig:Mag-Dirac-m}(a) (pure Dirac), we see a quantization in the slope at values of $\nu e/h$ (note: the Hall plateaus have been offset for clarity) where $\nu$ is a half-integer (beginning at +1/2 for positive $\mu$).  This corresponds to the half-integer quantum Hall effect: $\sigma_H=(e^2/h)\nu$, where $\nu=\pm 1/2, \pm 3/2,...$.  Note: for graphene, two-fold valley and spin degeneracies would be included which give rise to the well known filling factors $\pm 2, \pm 6, \pm 10,...$.  In the inset of Fig.~\ref{fig:Mag-Dirac-m}(b) (small $E_0$), we again see a half-integer quantization of the slope\cite{Wang:2010, ZLi:2014}; however, now it begins at $\nu=-1/2$.  This is due to the Schr{\"o}dinger term moving the location of the $N=0$ Landau level to positive energy.  Therefore, at energies below $E_0/2$, the highest filled level is $N=1$, $s=-$.  If we were to look at negative values of $\mu$, we would see the full negative half-integer quantization.  We note that the addition of a small Schr{\"o}dinger correction to the Hamiltonian does not break the half-integer quantization of the pure Dirac limit.  It does, however, result in a negative Hall conductivity for positive $\mu<E_0/2$.

To see this robust quantization, return to Eqn.~\eqref{Omega-TI}.  Taking the derivative with respect to $\mu$, we obtain
\begin{align}
\frac{\partial\tilde{\Omega}}{\partial\mu}=\frac{eB}{h}\left[\frac{1}{2}-\Theta(\mu-E_0/2)-\sum_{N=1}^\infty\Theta(\mu-E_{N,+})\right],
\end{align}
for $\mu>0$, where we ignore the $\delta$-functions resulting from the derivatives of the $\Theta$-functions.  Differentiating with respect to $B$, we obtain
\begin{align}
-\frac{\partial}{\partial B}\frac{\partial\tilde{\Omega}}{\partial\mu}\equiv\frac{\partial M}{\partial\mu}=\frac{e}{h}\left[-\frac{1}{2}+\Theta(\mu-E_0/2)+\sum_{N=1}^\infty\Theta(\mu-E_{N,+})\right].
\end{align}
It is clear, that for $\mu<E_0/2$, the Hall plateau occurs at -1/2, for $E_0/2<\mu<E_{1,+}$ the quantization is 1/2, etcetera.

\subsection{Comparison with the Schr{\"o}dinger Limit}

For comparison, it is useful to consider the dominant-Schr{\"o}dinger regime ($E_0\gg E_1$).  This limit is well understood, and will, therefore, not be discussed in detail.  In this system, the grand thermodynamic potential is\cite{Sharapov:2004}
\begin{align}\label{OmegaSFull}
\Omega_{\rm NR}(T,\mu)=-T\int_{-\infty}^{\infty} N(\omega) \rm{ln}\left(1+e^{(\mu-\omega)/T}\right)d\omega,
\end{align}
where, again, $N(\omega)$ is given by Eqn.~\eqref{DOS}.  At $T=0$, this gives
\begin{align}\label{OmegaS}
\Omega_{\rm NR}(0,\mu)&=-\frac{eB}{h}\bigg[(\mu-E_0/2)\Theta(\mu-E_0/2)\notag\\
&+\sum_{N=1, s=\pm}^{\infty}(\mu-E_{N,s})\Theta(\mu-E_{N,s})\bigg].
\end{align}
Again, the magnetization is given by $M=-\partial\Omega_{\rm NR}/\partial B$.

In the pure Schr{\"o}dinger limit ($E_1=0$), the Landau levels evolve linearly with $B$ unlike the $\sqrt{B}$ dependence observed for Dirac fermions [see Fig.~\ref{fig:LL-Dirac}(a)].  The $N=0,1,2,...$ levels for $s=+$ are degenerate in energy with the $N=1,2,3,..$ Landau levels for $s=-$.  There is no $N=0,\,s=-$ level.  The Hall conductivity is given by an even-integer quantization (i.e. $\sigma_H=(e^2/h)\nu$ where $\nu=0,2,4,...$).

\begin{figure}[h!]
\begin{center}
\includegraphics[width=0.9\linewidth]{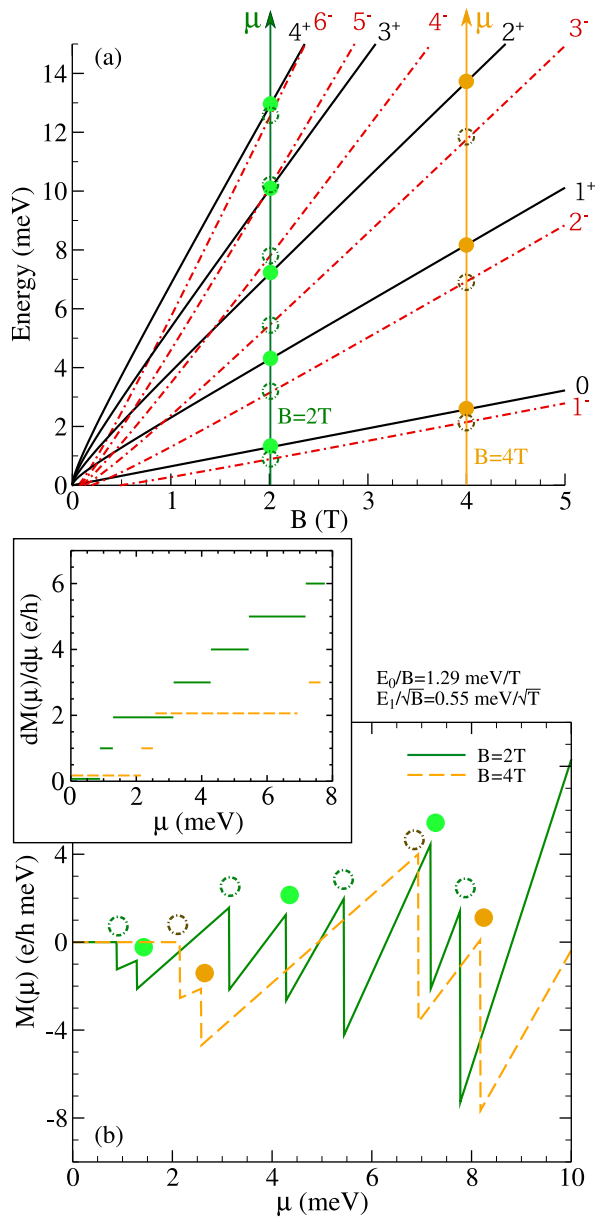}
\end{center}
\caption{\label{fig:Mag-Schro+D}(Color online) (a) Landau level dispersion for the spintronic regime ($E_0\gg E_1$).  The solid black and dashed red curves show the Landau level evolution with $B$.  The dashed and solid circles show the intersection of the Landau levels with the lines of constant $B$ used in (b). (b) Magnetization as a function of $\mu$. The results are shown for the two fixed values of $B$ corresponding to the vertical lines in (a).  The circles above the peaks illustrate the intersection of the constant $B$ lines with the Landau levels.  A double peak feature is seen due to a splitting of the Landau levels. The inset of (b) shows $\partial M(\mu)/\partial\mu=(1/e)\sigma_H$.  Here, $\sigma_H$ is quantized in integer values of $e^2/h$ due to the breaking of the Landau level degeneracy.  Some plateaus are displaced from their integer values for clarity.
}
\end{figure}
If a small relativistic contribution is included, the results of Fig.~\ref{fig:Mag-Schro+D}(a) and (b) are obtained for the Landau level dispersion as a function of $B$ and $M(\mu)$ as a function of $\mu$, respectively.  The Landau levels are no longer linear in $B$ but have an additional $\sqrt{B}$ dependence.  As a result, the degeneracy of the levels is broken.  The impact of the degeneracy breaking of the Landau levels on the magnetization is seen in Fig.~\ref{fig:Mag-Schro+D}(b).  For small $E_1$, the Landau level spectrum is close to the sum of two Schr{\"o}dinger systems which are separated by a small energy difference.  As a result, the peaks in the magnetization split in two but remain close in energy.  This affects the quantization of the Hall conductivity [see the inset of Fig.~\ref{fig:Mag-Schro+D}(b)].  With the breaking of the two-fold degeneracy, $\sigma_H$ is no longer quantized in even integers but can take any integer value\cite{ZLi:2014}.  For small $E_1$, the odd integer plateaus exist over a much smaller range of $\mu$ than the even-integer values.  As $E_1$ is increased, the odd integer steps become more prominent.  Unlike, the robust half-integer quantization seen in the relativistic limit (see the previous subsection), the addition of a relativistic term to the pure Schr{\"o}dinger regime changes the Hall conductivity from an even-integer to an integer quantization.  This quantization can be seen by returning to Eqn.~\eqref{OmegaS} and taking the derivative with respect to $\mu$, giving
\begin{align}
\frac{\partial\Omega_{\rm NR}}{\partial\mu}=-\frac{eB}{h}\left[\Theta(\mu-E_0/2)+\sum_{N=1, s=\pm}^\infty\Theta(\mu-E_{N,s})\right].
\end{align}
Taking the derivative with respect to $B$, we obtain
\begin{align}
-\frac{\partial}{\partial B}\frac{\partial\Omega_{\rm NR}}{\partial\mu}\equiv\frac{\partial M}{\partial\mu}=\frac{e}{h}\left[\Theta(\mu-E_0/2)+\sum_{N=1,s=\pm}^\infty\Theta(\mu-E_{N,s})\right].
\end{align}
For the pure Schr{\"o}dinger limit, $E_{N,-}$ is degenerate with $E_{N-1,+}$. For $\mu<E_0/2$, $\sigma_H=0$. The first plateau occurs for $E_0/2<\mu<E_{1,+}$ and has a value of 2 since $E_{1,-}=E_0/2$.  The next step occurs for $\mu>E_{1,+}$ which is degenerate with $E_{2,-}$ and so $\sigma_H$ is incremented by $2e^2/h$.  All further steps are also by two.  The inclusion of a small spin-orbit contribution breaks the degeneracy of the $s=\pm$ bands and now the steps occur at all integer values.  As the energy difference between $E_{N,-}$ and $E_{N-1,+}$ is small for $E_0\gg E_1$, the even integer steps are only visible over a small range of $\mu$.  For a more detailed discussion of the Hall effect in this system, the reader is referred to Ref.~\cite{ZLi:2014}.

\section{Magnetic Oscillations}

\subsection{Dirac Limit}

We now turn our attention to the purely oscillating part of the magnetization.  To begin, note that Eqn.~\eqref{DOS} can be written as
\begin{align}\label{DOS-dw}
N(\omega)=\frac{eB}{h}\frac{d}{d\omega}\left[\Theta(\omega-E_0/2)+\sum_{N=1, s=\pm}^{\infty}\Theta(\omega-E_{N,s})\right].
\end{align}
The oscillatory part of Eqn.~\eqref{DOS-dw} is obtained by applying the Poisson formula
\begin{align}
\sum_{N=1}^\infty F(N)&=-\frac{1}{2}F(0)+\int_0^\infty F(x)dx\notag\\
&+2\sum_{k=1}^\infty\int_0^\infty F(x)\rm{cos}(2\pi kx)dx.
\end{align}
This gives (see Eqn.~(19) in Ref.~\cite{Suprunenko:2008}),
\begin{align}\label{DOS-Osc}
N(\omega)&=\frac{eB}{h}\frac{d}{d\omega}\left\lbrace\frac{1}{2}\Theta(\omega-E_0/2)-\frac{1}{2}\Theta(\omega+E_0/2)\right.\notag\\
&+\left[\Theta(\omega+E_0/2)+\Theta(\omega-E_0/2)-\Theta(\omega-\omega_{\rm min})\right]\notag\\
&\times\left[x_1+\sum_{k=1}^\infty\frac{1}{\pi k}\rm{sin}(2\pi kx_1)\right]\notag\\
&+\left. \Theta(\omega-\omega_{\rm min})\left[x_2+\sum_{k=1}^\infty\frac{1}{\pi k}\rm{sin}(2\pi kx_2)\right]\right\rbrace,
\end{align}
where
\begin{align}
\omega_{\rm min}=-\frac{E_0P}{2}-\frac{E_0}{8P},
\end{align}
and
\begin{align}\label{xi}
x_i=P+\frac{\omega}{E_0}+(-)^i\sqrt{P^2+\frac{2P\omega}{E_0}+\frac{1}{4}}.
\end{align}

All the necessary information about the de-Haas-van-Alphen (dHvA) oscillations can be extracted from the first term of Eqn.~\eqref{Omega-int}.  Thus, we examine
\begin{align}
\bar{\Omega}(0,\mu)=\int_{0^+}^{\mu}(\omega-\mu)N(\omega)d\omega,
\end{align}
where the vacuum, and $eB\mu/(2h)$ terms are excluded as we are only interested in the oscillating part of $M(\mu)$.  In addition, only the trigonometric terms of $N(\omega)$ [Eqn.~\eqref{DOS-Osc}] will contribute to the oscillations.  As is argued by Suprunenko \emph{et al}\cite{Suprunenko:2008}, since we are working at low energy, we should only consider small values of $N$ and thus, in the relativistic regime, we should only include the $x_1$ term.

Focusing only on the terms which contribute to the oscillations, the magnetization is given by
\begin{align}\label{M-bar}
\bar{M}(0,\mu)=-\frac{d}{dB}\int_0^\mu(\omega-\mu)N_{\rm osc}(\omega)d\omega,
\end{align}
where
\begin{align}\label{DOS-Osc-part}
N_{\rm osc}(\omega)=\frac{eB}{h}\sum_{k=1}^\infty\frac{d}{d\omega}\bar{N}^k_{\rm osc}(\omega),
\end{align}
and
\begin{align}
\bar{N}^k_{\rm osc}(\omega)=&\left[\Theta(\omega+E_0/2)+\Theta(\omega -E_0/2)-\Theta(\omega-\omega_{\rm min})\right]\notag\\
&\times\frac{{\rm sin}(2\pi k x_1)}{\pi k}.
\end{align}
As we take only positive $\mu$ and $E_0$, this may be written as
\begin{align}
\bar{N}^k_{\rm osc}(\omega)=\Theta(\omega -E_0/2)\frac{{\rm sin}(2\pi k x_1)}{\pi k}.
\end{align}
In the relativistic limit, we use the approximation
\begin{align}
x_1\approx\frac{\omega^2}{2E_1^2}\left(1-\frac{\omega}{m v_F^2}\right)-\frac{E_0}{8mv_F^2},
\end{align}
which can be found by Taylor expanding Eqn.~\eqref{xi}.  In what follows, we will only keep amplitudes which are first order in $1/m$.  Substituting the $k$-th component of Eqn.~\eqref{DOS-Osc-part} into Eqn.~\eqref{M-bar}, and integrating by parts, we obtain
\begin{align}
\bar{M}_k(0,\mu)\approx\frac{d}{dB}\left\lbrace\frac{eB}{h}\int_0^\mu \bar{N}^k_{\rm osc}(\omega)d\omega\right\rbrace.
\end{align}
Therefore, assuming $\mu>E_0/2$, we have
\begin{align}
\bar{M}_k(0,\mu)\approx\frac{d}{dB}&\left\lbrace\frac{eB}{\pi kh}\left[\int_0^\mu{\rm sin}(2\pi kx_1)d\omega\right.\right.\notag\\
&\left.\left.-\int_0^{\frac{E_0}{2}}{\rm sin}(2\pi kx_1)d\omega\right]\right\rbrace.
\end{align}
We ignore the second term as, for the limit of interest ($B\rightarrow 0$), $E_0\rightarrow 0$; hence, expanding the sine to lowest order and performing the integral, gives a magnetization that goes like $B^2$ and is thus negligible.

We are left with evaluating
\begin{align}
M^k_{\rm osc}(0,\mu)=\frac{e}{\pi kh}\int_0^\mu\left[{\rm sin}(2\pi kx_1)-2\pi kx_1{\rm cos}(2\pi kx_1)\right]d\omega.
\end{align}
To solve this integral, define
\begin{align}
y\equiv \omega^2\left[1-\frac{\omega}{mv_F^2}\right]-\frac{E_0^2}{4}.
\end{align}
For $m\rightarrow\infty$, $y\approx\omega^2$ so $\sqrt{y}\approx\omega$.  Therefore, we can write
\begin{align}
y\approx\omega^2\left[1-\frac{\sqrt{y}}{mv_F^2}\right]-\frac{E_0^2}{4}.
\end{align}
Our simplified integral is now
\begin{align}
 M^k_{\rm osc}(0,\mu)&\approx\frac{e}{2\pi kh}\int_0^{\alpha}\left[{\rm sin}\left(\frac{\pi ky}{E_1^2}\right)-\frac{\pi ky}{E_1^2}{\rm cos}\left(\frac{\pi ky}{E_1^2}\right)\right]\notag\\
 &\times\left[1+\frac{\sqrt{y}}{mv_F^2}\right]\frac{dy}{\sqrt{y}}.
\end{align}
where
\begin{align}
\alpha=\mu^2\left[1-\frac{\mu}{mv_F^2}\right]-\frac{E_0^2}{4}.
\end{align}
We then make the substitution $y\equiv x\alpha$, to obtain
\begin{align}\label{Mag-osc-int}
 M^k_{\rm osc}(0,\mu)&\approx\frac{e}{2\pi kh}\int_0^1\left\lbrace\sqrt{\frac{\mu^2[1-\mu/(mv_F^2)]}{x}}\right.\notag\\
 &\times[{\rm sin}(ax)-ax{\rm cos}(ax)]\notag\\
&\left.+\frac{\mu^2}{mv_F^2}[{\rm sin}(ax)-ax{\rm cos}(ax)]\right\rbrace dx,
\end{align}
where $a\equiv\pi k \alpha/E_1^2$.

Let us consider the first integral of Eqn.~\eqref{Mag-osc-int}.  We have
\begin{align}
I_1&=\int_0^1[{\rm sin}(ax)-ax{\rm cos}(ax)]\frac{dx}{\sqrt{x}}\cr
&=\int_0^1{\rm sin}(ax)\frac{dx}{\sqrt{x}}-\int_0^1\sqrt{x}\frac{d}{dx}{\rm sin}(ax)dx.
\end{align}
This can be integrated by parts to give
\begin{align}
I_1=3\sqrt{\frac{\pi}{2a}}\mathscr{S}\left(\sqrt{\frac{2a}{\pi}}\right)-{\rm sin}a,
\end{align}
where we have used the definition of the Fresnel sine integral
\begin{align}
\mathscr{S}(z)=\int_0^z{\rm sin}\left(\frac{1}{2}\pi t^2\right)dt.
\end{align}
We are interested in the oscillations for small $B$.  In the limit $B\rightarrow 0$, $a\rightarrow\infty$.  Expanding $\mathscr{S}(\sqrt{2a/\pi})$ to first order in $1/a$, we obtain
\begin{align}
\mathscr{S}\left(\sqrt{\frac{2a}{\pi}}\right)\approx\frac{1}{2}-\frac{1}{\sqrt{2\pi a}}{\rm cos}a;
\end{align}
thus, in the limit of interest,
\begin{align}
I_1\approx-{\rm sin}a.
\end{align}

Next, consider the second term of Eqn.~\eqref{Mag-osc-int}.  This is a standard integral and gives
\begin{align}
I_2=\int_0^1[{\rm sin}(ax)-ax{\rm cos}(ax)]dx\approx -{\rm sin}a
\end{align}
in the limit of $B\rightarrow 0$.  Combining $I_1$ and $I_2$, we obtain
\begin{align}\label{MOSC}
M_{\rm osc}(0,\mu)\approx -\frac{e}{h}\frac{\mu }{2}\sum_{k=1}^\infty\left[1+\frac{\mu }{2mv_F^2}\right]\frac{{\rm sin}(2\pi k x_1)}{\pi k},
\end{align}
where
\begin{align}\label{x1-D}
x_1\approx \frac{\mu^2}{2\hbar v_F^2eB}\left(1-\frac{\mu }{mv_F^2}\right)-\frac{\hbar eB}{8m^2v_F^2}.
\end{align}
If we compare this to the customary\cite{Luk:2004}
\begin{align}
x_1=\frac{\hbar A(\mu)}{2\pi eB}-\gamma,
\end{align}
the coefficient of the $1/B$ dependence in Eqn.~\eqref{x1-D} is indeed the area of the cyclotron orbit:
\begin{align}\label{AreaD}
A(\mu)\approx\frac{\pi\mu^2}{\hbar^2 v_F^2}\left(1-\frac{\mu}{mv_F^2}\right).
\end{align}
The remainder is a phase shift which has a linear dependence on $B$. It is new and is not part of a standard semiclassical quantization scheme\cite{Fuchs:2010}.  It is
\begin{align}\label{gammaD}
\gamma=\frac{\hbar eB}{8m^2v_F^2}.
\end{align}
In the pure Dirac limit ($m\rightarrow\infty$), $A(\mu)$ reduces to the correct value\cite{Luk:2004,Suprunenko:2008} of $\pi\mu^2/(\hbar^2 v_F^2)$.  The inclusion of a small Schr{\"o}dinger contribution gives a correction of $-[\pi\mu^2/(\hbar^2 v_F^2)][\mu/(mv_F^2)]$ to $A(\mu)$ and the cyclotron orbit area is reduced due to the narrowing of the conduction band [see Fig.~\ref{fig:Energy}(a)].  The pure Dirac limit should also have a phase shift of 0 associated with a Berry's phase of $\pi$\cite{Suprunenko:2008}.  Here, we find the inclusion of an additional $E_0$ term results in a finite phase shift of $\hbar eB/(m^2v_F^2)$ which is linear in $B$ but very small for $m\rightarrow\infty$.  There is also a correction to the overall amplitude of the quantum oscillations.  While there can be a significant Schr{\"o}dinger contribution to the low-energy Hamiltonian of a topological insulator, we find here [Eqn.~\eqref{gammaD}] that the phase offset $\gamma$ of the magnetic oscillations is essentially zero; this is in agreement with previous semiclassical considerations\cite{Wright:2013, Wright:2013a, Fuchs:2010, Taskin:2011}.  This is the same  result as for pure relativistic particles for which the Berry phase is $\pi$.  The result is in agreement with experimental findings.  For example, the high-field SdH data in Bi$_2$Te$_2$Se by Xiong \emph{et al}\cite{Xiong:2012} shows zero offset.  The same holds for many other topological insulators as documented in the extensive review by Ando\cite{Ando:2013}.  This signature of zero offset has often been used to distinguish between oscillations coming from the surface states and those originating from the bulk\cite{Ando:2013}.  In SdH (oscillations in the conductivity) and dHvA (magnetization oscillations), experiment can also extract the cyclotron orbit area\cite{Lawson:2012}.  Equation~\eqref{AreaD} shows that this quantity contains a correction from the pure Dirac result of order $\mu/(mv_F^2)$ which may allow one to extract the Schr{\"o}dinger mass $m$ from such data.

\subsection{Dingle Factor}\label{sec:Dingle}

The Landau level broadening due to impurity scattering can be included by convolving the grand thermodynamic potential with a scattering function $P_\Gamma(\varepsilon)$.  That is\cite{Sharapov:2004},
\begin{align}\label{Omega-Imp}
\Omega(\mu)=\int_{-\infty}^\infty d\omega P_{\Gamma}(\omega-\mu)\Omega(\omega),
\end{align}
where
\begin{align}
P_\Gamma(\varepsilon)=\frac{\Gamma}{\pi\left(\varepsilon^2+\Gamma^2\right)},
\end{align}
and $\Gamma$ is a small broadening parameter.  

To solve for the Dingle factor, consider the magnetization in the presence of impurities:
\begin{align}\label{MOSC-Dingle}
M_{\rm osc}(\mu,\Gamma)=\int_{-\infty}^\infty d\omega \frac{\Gamma}{\pi\left[(\omega-\mu)^2+\Gamma^2\right]}M_{\rm osc}(\omega),
\end{align}
where $M_{\rm osc}(\omega)$ is given by letting $\mu\rightarrow\omega$ in Eqn.~\eqref{MOSC} for the relativistic regime.  To obtain a first order approximation to the Dingle factor, we note that the Lorentzian is peaked around $\omega=\mu$ and that $\omega/(2mv_F^2)$ is small.  We, therefore, fix $\omega$ at $\mu$ in these terms.  Here, we also assume $\Gamma\ll\mu$.  The $k^{\rm th}$ component of Eqn.~\eqref{MOSC-Dingle} becomes
\begin{align}
M^k_{\rm osc}(\mu,\Gamma)&\approx\int_{-\infty}^\infty d\omega \frac{\Gamma}{\pi\left[(\omega-\mu)^2+\Gamma^2\right]}\left\lbrace-\frac{e}{h}\frac{\omega}{2\pi k}\left(1+\frac{\mu }{mv_F^2}\right)\right.\cr
&\left.\times{\rm sin}\left[\frac{k\pi}{E_1^2}\omega^2\left(1-\frac{\mu }{mv_F^2}\right)-\frac{\hbar eB}{8m^2v_F^2}\right]\right\rbrace.
\end{align}
If we ignore the phase factor for the purpose of integrating, and define $1/E_1^{*2}\equiv(1/E_1^2)[1-\mu/(mv_F^2)]$, we have mapped our results onto the pure Dirac limit and obtain the renormalized results of Ref.~\cite{Sharapov:2004} [see their Eqn.~(8.10)]:
\begin{align}
M(\mu,\Gamma)\approx -\frac{e}{h}\frac{\mu^2-\Gamma^2}{2\mu}\left(1+\frac{\mu }{2mv_F^2}\right)\sum_{k=1}^\infty\frac{{\rm sin}(2\pi k\tilde{x}_1)}{\pi k}R_D,
\end{align}
where
\begin{align}
\tilde{x}_1\approx\frac{\mu^2-\Gamma^2}{2\hbar v_F^2eB}\left(1-\frac{\mu }{mv_F^2}\right)-\frac{\hbar eB}{8m^2v_F^2},
\end{align}
and
\begin{align}\label{Dingle-Dirac}
R_{D}=e^{\displaystyle -2\pi\Gamma k\frac{\mu}{\hbar v_F^2eB}\left(1-\frac{\mu}{mv_F^2}\right)}.
\end{align}
This reduces to the expected\cite{Sharapov:2004} $R_{D}=$exp$[-2\pi\Gamma\mu/(\hbar v_F^2eB)]$ in the pure-Dirac limit. We note that in this regime, the damping is dependent on the chemical potential\cite{Sharapov:2004}; thus, for large $\mu$, magnetic oscillations would be difficult to observe\cite{Sharapov:2004}.  When particle-hole symmetry is broken by a subdominant Schr{\"o}dinger piece, the damping in the Dingle factor $R_D$ for a given $\mu$ is reduced by a factor of $1-\mu/(mv_F^2)$ which makes it more favourable for observation of magnetic oscillations as compared to the pure relativistic case.

\subsection{Finite T}

Finite temperature effects are included by convolving $M_{\rm osc}(\omega)$ with $P_T(\omega-\mu)$, where\cite{Sharapov:2004}
\begin{align}\label{Omega-T}
P_T(\varepsilon)=\frac{1}{\displaystyle 4T\rm{cosh}^2\left(\frac{\varepsilon}{2T}\right)},
\end{align}
and, $M_{\rm osc}(\omega)$ is the result of letting $\mu\rightarrow\omega$ in Eqn.~\eqref{MOSC}.  Again, the damping factor is peaked around $\omega=\mu$.  By applying the same procedure as in Sec.~\ref{sec:Dingle}, we map our results onto those of Ref.~\cite{Sharapov:2004} [see their Eqns.~(8.16) and (8.17)] and obtain
\begin{align}
M(\mu,T)\approx -\frac{e}{h}\frac{\mu}{2}\left(1+\frac{\mu }{2mv_F^2}\right)\sum_{k=1}^\infty\frac{{\rm sin}(2\pi k x_1)}{\pi k}R_T,
\end{align}
where
\begin{align}\label{Dingle-T}
R_T=\frac{k\lambda}{{\rm sinh}(k\lambda)},
\end{align}
with
\begin{align}
\lambda=\frac{2\pi^2T\mu}{\hbar v_F^2eB}\left(1-\frac{\mu }{mv_F^2}\right).
\end{align}
We note that $R_T$ is $1$ in the limit $T\rightarrow 0$ as it must be.  Like impurity scattering, the damping of oscillations due to temperature is dependent on $\mu$.  For large $\mu$, magnetic oscillations are difficult to observe.  The introduction of a small Schr\"odinger term reduces the damping by $1-\mu/(mv_F^2)$ similar to what we found in the previous section for impurity scattering.

\section{Conclusions}

We have computed the magnetization of the topologically protected helical electronic states that exist at the surface of a three dimensional topological insulator.  We focus on adding a small Schr{\"o}dinger quadratic-in-momentum term to a dominant linear-in-momentum Dirac term. Typical for a topological insulator, is a Schr{\"o}dinger mass ($m$) on the order of $1/10$ the bare electron mass and a Dirac Fermi velocity ($v_F$) of order $\sim 4\times 10^5$m/s.  Thus, the relevant magnetic energy scales for $B=1$T are $E_0\sim 1.0$meV and $E_1$ of order 10.0meV.  The ratio of Schr{\"o}dinger to Dirac energy scales $E_0/E_1$ is much less than one.  This is the opposite limit to that found for semiconductors used in spintronic applications where the spin-orbit coupling is 100 times smaller and $E_0/E_1$ is much greater than one.   

In the topological insulator limit, the magnetization as a function of chemical potential $\mu$ shows the usual jagged sawtooth oscillations.  The teeth occur at values of $\mu$ which reflect the underlying Landau level energies.  While the distance between the teeth for fixed $B$ as a function of $\mu$ is altered for a small Schr{\"o}dinger contribution, there is no qualitative difference to the pure Dirac limit.  For comparison, and in sharp contrast to this finding, adding a small spin-orbit coupling term to a quadratic band introduces a new distinct set of teeth to those that arise when $v_F=0$.  These new teeth are mush less prominent and appear grafted on the side of and slightly displaced from the main set.  They disappear when spin-orbit coupling is neglected and their distinctness increases with $v_F$.

The slope of the magnetization as a function of $\mu$ reflects the quantization of the Hall conductivity and is found to remain unchanged for a topological insulator as the Schr{\"o}dinger contribution is increased and the half integer quantization of the pure Dirac regime is retained.  Of course, the amount by which the chemical potential needs to be incremented to go from one plateau to the next is changed.  These results are in striking contrast to the case when the spin-orbit term is small and the Schr{\"o}dinger mass term dominates.  While the spin-orbit coupling splits the Landau level degeneracy of the pure non-relativistic case, the well known integer quantization of the plateaus remains even though two distinct sets are observed.  The second set merges with the dominant set for $v_F=0$, and increases in prominence with increasing spin-orbit coupling.

Applying the Poisson formula to the thermodynamic potential, we extract an analytic expression for the magnetic oscillations $M_{\rm osc}$ which includes a first correction in a small Schr{\"o}dinger term.  Our new expression properly reduces to the known result of Ref.~\cite{Sharapov:2004} when we set $E_0$ to zero which is the relativistic generalization of the classical Lifshitz-Kosevich (LZ) result.  In this case, the phase offset which appears in the semiclassical expression for $M_{\rm osc}$ is equal to zero in contrast to the value of 1/2 of LK theory\cite{Luk:2004}.  This difference can be traced back to a Berry phase which is $\pi$ in the Dirac limit and zero in the Schr{\"o}dinger limit.  When a sub-dominant mass term is added to the Dirac limit, the phase offset is no longer zero but has a correction of order $(E_0/E_1)^2$ which is equal to $\hbar eB/(8m^2v_F^2)$. This new contribution to the phase is zero in the pure relativistic case $m\rightarrow\infty$ and is also negligible when the magnetic field tends to zero.  At the same time, the Berry phase is totally unchanged by the $E_0/E_1$ correction term and retains its value of $\pi$.  Nevertheless, in a topological insulator, there is a small phase offset in the quantum oscillations which has its origin in the bending of the electronic dispersion curves away from linearity which narrows slightly the  cross-section of the conduction band Dirac cone.

We have considered the effect of impurity scattering on the magnetization in a constant scattering rate $(\Gamma)$ approximation.  We assume the same value of $\Gamma$ applies to all the Landau levels.  This provides a Dingle factor in $M_{\rm osc}$.  In the pure-relativistic case, the exponential argument of the damping factor depends linearly on chemical potential\cite{Sharapov:2004}.  Here, we find a correction to this damping by a further multiplicative factor in the exponential of $1-\mu/(mv_F^2)$.  This reduces the effectiveness of damping over its pure relativistic value. A similar correction to the temperature factor is also noted.

\begin{acknowledgments}
We thank E. J. Nicol for discussions.  This work has been supported by the Natural Science and Engineering Research Council of Canada and, in part, by the Canadian Institute for Advanced Research.
\end{acknowledgments}

\bibliographystyle{apsrev4-1}
\bibliography{MO-arxiv}

\end{document}